\documentstyle[12pt]{article}
\font\bigbf=cmbx10 scaled\magstep3

\begin{document}
\begin{center}
 {\bigbf Quantum gravity effects in the CGHS model of collapse 
   to a black hole}\\

\vspace*{0.35in}

\large

Madhavan Varadarajan

\normalsize

{\sl Raman Research Institute, Bangalore 560 080, India.}
\\
madhavan@rri.ernet.in\\
\vspace{.5in}
October 3, 1997\\
\vspace{.5in}
ABSTRACT
We show that only a sector of the classical solution space of the 
CGHS model describes formation of black holes through collapse 
of matter. This sector has either right or left moving matter. 
We describe the sector which has left moving matter in canonical language.
In the nonperturbative quantum theory all operators are expressed in terms of 
the matter field operator which is represented on a Fock space. 
 We discuss existence of 
large quantum fluctuations of the metric operator when the matter field is 
approximately classical.  We end with some comments 
which may pertain to Hawking radiation in the context of the model.

\end{center}

\pagebreak

\setcounter{page}{1}

\section{Introduction}

We regard the CGHS model \cite{cghs} as a 2d classical field theory in its 
own right rather than as string inspired. 
We view the dilaton field as just another classical field 
rather than as describing  the string coupling.
The theory  
shares two important features with 4d general relativity
 - it is a nonlinear  diffeomorphism invariant field theory 
whose solutions describe spacetime metrics and {\em some} of these
solutions correspond to black hole formation through matter 
collapse.  
Since the  model is classically exactly solvable and 
modulo certain very
important qualifications, has been (non-perturbatively)
canonically quantized via the Dirac procedure \cite{jkm},
we use it as a toy model for these features of 4d quantum 
general relativity.

In this work we prove some results regarding the properties of classical
 solutions to the model using a general relativist's point of view. Next, we 
discuss some quantum properties of the spacetime geometry along the lines
of \cite{abhay}. We end with a few speculative remarks concerning Hawking 
radiation in the model. The discussion and analysis of the quantum mechanics
of the model is based on the recent work \cite{jkm}. 

The outline of the paper is as follows.
In order to interpret the quantum theory of \cite{jkm} it is 
essential to understand
the classical solution space \cite{hayward}.
In section II, we
show that `most' classical solutions {\em do not} correspond to matter 
collapsing to form a black hole. More specifically, we show that
if both left and right moving matter is present, the spacetime does  not 
represent black hole formation through matter 
collapse.\footnote{A similar result is 
asserted in section 8 of \cite{hayward}.}
 However, if only `one sided' 
matter is present, it is possible to obtain solutions describing 
collapse to a black hole.

In fact, without the restriction to one sided collapse, 
it is difficult to characterise broad properties of the spacetime 
in terms of properties of the matter field distribution (see, however 
\cite{hayward}). We have very 
little control over the solution space and do not understand exactly
what facets of 4d general relativistic physics, if any, are modelled by the 
solutions. In order to retain the solutions corresponding to the 
collapsing black hole spacetimes  as well as to have better control on the 
space of solutions, we restrict attention to the one sided collapse sector
in the remainder of the paper.


Solutions to the CGHS model are most simply described in Kruskal
like null cone coordinates $X^{\pm}$ \cite{jkm}. The one sided 
collapse solutions which we analyse describe a single 
black hole  spacetime  and correspond to the restriction 
 $X^{\pm} >0$.
 These solutions can be analytically 
extended through the entire $X^{\pm}$ plane (indeed, the quantum theory of 
\cite{jkm} seems to require consideration of such extensions!).
We show, through Penrose diagrams, how the physical spacetime is embedded in, 
and analytically extended to, the full $X^{\pm}$
plane. This completes our analysis of the classical solution 
space. 

In section III, we turn to the Hamiltonian description of 
the model. Since we are interested in the 1 sided collapse situation, 
we restrict the description in \cite{jkm} suitably, 
by setting the left mass 
of the spacetime and the right moving matter 
modes to zero. We adapt the quantization of \cite{jkm} 
to the 1 sided collapse case. To make contact with the 
semiclassical treatment of Hawking radiation in the literature 
(see, for example, 
\cite{strominger}), left {\em and} right moving modes are needed.
Since one set of modes have been frozen in our analysis, 
we do not discuss Hawking radiation related issues except for 
some comments at the end of the paper in section V. 
Instead, we focus on 
other issues arising in quantum gravity. 
We calculate quantum fluctuations of the metric operator when the matter 
fields
are approximately classical (metric operator fluctuations have been discussed
earlier in \cite{mikovic} and, in the context of  spacetimes with an 
internal boundary, in \cite{fluct} ).
We show
that large quantum gravity effects as in the case of cylindrical
 waves \cite{abhay} are manifested even far away from the 
singularity (although not at spatial infinity).

The discussion of this section pertains to the quantum 
version of the analytic
extension to the entire $X^{\pm}$ plane, of the 1 sided collapse 
solutions. In contrast, in section IV, we deal with (the 
canonical classical and quantum theory of) only the 
physical spacetime region $X^{\pm} >0$. To do this we 
appropriately modify the analysis of asymptotics in \cite{jkm}. 
The most 
direct route to the quantum theory is to first gauge fix (a la 
Mikovic \cite{mikovic}) and then quantize the resulting 
description. 
We obtain a Fock space representation based on the time choice 
$\ln({X^+ \over X^-})$ in contrast to the Fock space of section III
which was based on the time choice ${X^+ + X^- \over 2}$.
We repeat the analysis of section III regarding large
quantum gravity effects. In the process we find that the
operator corresponding to the spacetime metric at large values of $X^+$
cannot be represented on the Fock space of the quantum
theory.The implication is that the most natural representation
(which we have chosen) for the quantum theory may not be the correct one. 
We leave this as an open problem.

Section V contains concluding remarks including some comments on 
Hawking radiation in the context of the model. 

We do not attempt to review the vast amount of pertinent literature but 
instead refer the reader to review articles such as \cite{strominger}.

\subsection*{Notation} 
Besides standard conventions, we will use the following notation 
(from \cite{jkm}) throughout this paper: 
In the double null coordinates $X^\alpha=(X^+,X^-)$, many quantities 
depend only on $X^+$ or $X^-$, but not on both variables.
We will emphasize this by using only $X^+$ or $X^-$ as an argument of 
that function or functional.
For example, while $f(X)$ means that $f$ is a function of both $X^+$ 
and $X^-$, $f_{,+}(X^+)$ and $f_{,-}(X^-)$ mean that the derivatives
$f_{,+}$ and $f_{,-}$ depend only on $X^+$ and $X^-$, respectively.
Moreover, $f_{,\pm}(X^\pm)$ will serve as a shorthand notation to 
denote the function dependence of both $f_{,+}$ and $f_{,-}$ simultaneously.
${\cal I}_L^-,{\cal I}_R^-,{\cal I}_L^+$ and $ {\cal I}_R^+$
denote past left, past right, future left and  future right null
infinity respectively.

\section{Analysis of the classical solution space}

\subsection{The Action and the solution to the field equations}
We briefly recall the action and the solution to the field equations
for the CGHS model in the notation of \cite{jkm} (for details see \cite{jkm}).
In units in which the velocity of light, $c$, and the gravitational constant,
$G$, are unity, the action is 
\begin{equation}
S[y,\gamma_{\alpha\beta},f]={1\over 2}\int d^2\!X\ |\gamma|^{{1\over 2}}
\left(y R[\gamma]+4\kappa^2-\gamma^{\alpha\beta}f_{,\alpha}f_{,\beta}\right)
\ .
\label{eq:S}
\end{equation}
Here $y$ is the dilaton field, $\gamma_{\alpha\beta}$ is the spacetime 
metric (signature $(-+)$), and $f$ is a conformally coupled scalar field.
$R[\gamma]$ denotes the scalar curvature of $\gamma_{\alpha\beta}$, and 
$\kappa$ is a positive definite constant having the dimensions of inverse 
length.

To interpret the theory, we will treat $\gamma_{\alpha\beta}$ as an 
auxiliary metric and 
\begin{equation}
\bar\gamma_{\alpha\beta}:=y^{-1}\gamma_{\alpha\beta}
\label{eq:physical_metric}
\end{equation}
as the physical ``black hole'' metric. Since $y$ is a conformal factor, it 
 is restricted to be positive. However, note that 
since the field equations and 
(\ref{eq:S}) are well defined for $y \leq 0$, solutions with positive $y$ 
admit analytic extensions to $y\leq 0$. 
The solution to the field equations arising from (\ref{eq:S}) is as 
follows (for details see \cite{jkm}). $\gamma_{\alpha \beta}$ is flat. 
The remaining fields are most 
elegantly described in terms of the double null coordinates $X^{\pm}=Z\pm T$,
where $(Z,T)$ are the Minkowskian coordinates associated with the flat 
auxilliary metric. Then 
the spacetime line element associated with the metric $\gamma_{\alpha \beta}$
is  
\begin{equation}
ds^2=dX^+ dX^-\ ,
\label{eq:flatmetric}
\end{equation}
the matter field is the sum of left and right movers 
\begin{equation}
f(X)=f_+(X^+)+f_-(X^-)\ ,
\end{equation}
and in the conformal gauge \cite{cghs}
\begin{equation}
y(X)=\kappa^2 X^+ X^- + y_+(X^+) + y_-(X^-)\ .
\label{eq:y1}
\end{equation}
Here
\begin{equation}
y_\pm(X^\pm)=-\int^{X^\pm} d\bar X{}^\pm\int^{\bar X{}^\pm}
d\bar{\bar X}^\pm\left(f_{,\pm}(\bar{\bar X}^\pm)\right)^2 \ .
\label{eq:y2}
\end{equation}
Finally, the line element corresponding to the physical metric, 
$\bar\gamma_{\alpha \beta}$ is 
\begin{equation}
d{\bar s}^2={dX^+ dX^- \over y} \ .
\label{eq:dbars}
\end{equation}
Its scalar curvature is 
\begin{equation}
{\bar R} = 4 \left( {y_{,+-}\over y} - {y_{,+}y_{,-}\over y^2} \right) \ .
\end{equation}
For smooth matter fields, it is easy to see that curvature singularities
can occur only when  $y=0$ or $y=\infty$ (the converse may not be true).

\subsection{Unphysical nature of solutions with `both sided matter'}
We now analyse the physical spacetime structure corresponding to the
solutions described above.
We are interested in those solutions which describe matter collapse to a black
 hole. So for spacetimes of physical interest we require that \\
\noindent {\bf (i)} A notion of (left past and future, right past and future) 
null infinities exists such that 
any light ray originating within the physical spacetime, 
traversing a region of no curvature singularities and reaching null infinity
should exhaust infinite affine parameter to do so. Further, null infinity 
is the locus of all such points. Each of left past, right past, left future 
and right future null infinity is a null surface 
diffeomorphic to the real line and forms part of the boundary of the 
spacetime. \\
\noindent {\bf (ii)} Only future singularities should exist.
Note that since $y$ is a conformal factor (\ref{eq:dbars}), it is required
to be positive. Any region {\em within} the physical spacetime 
where $y\leq 0$ is defined to be singular.

\vspace{3mm}

For simplicity we restrict the spacetime topology to be $R^2$. We also 
assume that the matter fields be of compact support at past null infinity.

Since the null infinities are null boundaries of the spacetime, they are 
labelled by lines of constant  $X^\pm$ (the constant could be finite or 
infinite). Thus, the physical spacetime is a subset of the entire
Minkowskian plane framed by boundaries made up of lines of constant $X^\pm$
and a future singularity. With this picture in mind let us further analyse 
the consequences  of {\bf (i)} and {\bf (ii)}. 

Further consideration of {\bf (i)} results in the following lemma.\\

\noindent {\bf Lemma:} If a section of null infinity is labelled by 
$X^+ =0$ or $X^- =0$ then {\bf (i)} implies that $y=0$ on this  section.

\noindent {\bf Proof:} Let a section of (right future or left past) null 
infinity be labelled by $X^+ =0$. Approach $X^+ =0$, through a nonsingular
 region  along an 
$X^-=$ constant $= a^-$  line  from $X^+= a^+$ ($a^+$ is finite). Let the 
normal  to this line be $k_a = \alpha (dX^{-})_a$.
For this line to be a geodesic ${\partial \alpha\over \partial X^{+}} =0$.
Choose $\alpha =1$. Let the affine parameter along this geodesic be 
$\lambda$. {\bf (i)} implies
\begin{equation}
|\lambda (X^-=a^-, X^+ =0) - \lambda (a^-, a^+)|
          = |\int_{a^+}^{0}{dX^+ \over \alpha y}|= \infty . 
\end{equation}
From {\bf (i)}, if $X^+=0$ is to label null infinity, 
 $y\rightarrow 0$ as $X^+ \rightarrow 0$ in such a way as to make the 
integral diverge.
Hence $y(a^-, 0) = 0$.

\vspace{5mm}

We now show that {\bf (i)} or 
{\bf (ii)} is violated if both left and right moving matter 
is present. For this, we examine (\ref{eq:y1}), (\ref{eq:y2}) and 
choose the lower limits 
of integration in (\ref{eq:y2}) as follows. Let the least value of $X^-$ be
$X^{-}_{0}$ on left future null infinity and that of $X^+$ be $X^{+}_{0}$ 
on right past null infinity. Then we specify $y$ as
\begin{equation}
y(X)=\kappa^2 X^+ X^- + y_+(X^+) + y_-(X^-) + a_+X^+ + a_-X^- + b .
\label{eq:y1l}
\end{equation}
where $a_{\pm}, b$ are constants and 
\begin{equation}
y_\pm(X^\pm)=-\int^{X^\pm}_{X^{\pm}_0} d\bar X{}^\pm
                    \int^{\bar X{}^\pm}_{X^{\pm}_0}
d\bar{\bar X}^\pm\left(f_{,\pm}(\bar{\bar X}^\pm)\right)^2
   \ .
\label{eq:y2l}
\end{equation}
The auxilliary flat metric determines $X^{\pm}$ only upto Poincare 
transformations. In this section, if $X^+_0$ ($X^-_0$) happens to be finite,
we use the translational freedom in $X^+$ ($X^-$) 
to set $X^+_0=0$ ($X^-_0=0$). 

Our strategy will be to 
 demand  {\bf (i)} or {\bf (ii)} and use the lemma
for exhaustive choices of ranges of $X^{\pm}$. Thus we {\em assume} that 
the physical spacetime satifies {\bf (i), (ii)} and that the boundaries of 
these ranges label the infinities of the spacetime. 
Singularities will occur 
inside the ranges when $y<0$ :\\
\noindent {\bf (A)}$-\infty < X^{\pm} < \infty$ : Past timelike infinity is 
labelled
by $(X^-, X^+) = (\infty, -\infty)$. As we approach this point, the first term
on the right hand side of (\ref{eq:y1l}) becomes arbitrarily negative 
and since it dominates the behaviour of $y$, it drives $y$ to  negative
values. 
The region $y<0$ must `intersect' past left and past right infinity.
Since $y$ cannot be negative, there must be a past singularity in the 
spacetime. Thus {\bf (ii)} rules out this range for $X^\pm$. 

\vspace{3mm}

\noindent {\bf (B)}$-\infty <  X^- < \infty \; , 0 < X^+ < \infty$  :
Left past null infinity is labelled by $X^+=0$. By the Lemma, $y(X^-, 0)=0$.
From (\ref{eq:y1l}) this gives, on left past null infinity 
\begin{equation}
y^-(X^-) + a_- X^-  + b= 0
\end{equation}
Differentiating this equation with respect $X^-$ yields 
$f_{,-}=0$. Thus {\bf (i)}
implies that there cannot be right moving matter for this choice of range.

All other choices of range can be handled by using the arguments in 
{\bf (A),(B)}. The conclusion is that either there is a past singularity in 
the
spacetime so that the corresponding range is ruled out or 
 $f_{,-}$ or 
$f_{,+}$  vanish. Thus we have proved the following statement:

{\em In the conformal gauge, if (i) and (ii) hold, then either left moving 
or right moving matter must vanish.}

Note that we have {\em not} proved the converse of this statement.

\subsection{One sided collapse to a black hole}
Having established that classical solutions of physical interest contain only
`one sided' matter, we turn to the analysis of (\ref{eq:y1l}) 
with $f^-=0$
(a similar analysis can be done for $f^+=0$).

We first identify the region of the $X^\pm$ plane corresponding to the
 physical spacetime. Let us fix the translation freedom in $X^\pm$ by setting 
$a_{\pm} =0 $ in (\ref{eq:y1l})
\footnote{This is a different choice from that used in Section 2.2}. 
Using arguments similar to those in the Lemma, {\bf (A)} and {\bf (B)}, 
it can be shown
that the only possible labellings of past null infinity which do not 
contradict
{\bf (i)}, {\bf (ii)} and $f_{,+} \neq 0$ are
 past left null infinity at 
$X^+= 0$ and past right null infinity at $X^- = \infty$.


Thus the solution of interest for the rest of the paper is 
\begin{equation}
y(X)=\kappa^2 X^+ X^- + y_+(X^+) \ .
\label{eq:y1l+}
\end{equation}
with 
\begin{equation}
y_+(X^+)=-\int^{X^+}_0 d\bar X{}^+
                    \int^{\bar X{}^+}_0
d\bar{\bar X}^+\left(f_{,+}(\bar{\bar X}^+)\right)^2
   \  ,
\label{eq:y2l+}
\end{equation}
and $X^+ > 0$. Let the support of $f^+$ be $\alpha < X^+ < \beta$. Note that 
within the physical spacetime $X^- \geq 0$ otherwise $y(x)$ can become 
negative. Note that  
\begin{equation}
y(x)=\kappa^2 X^+ X^- \;\; {\rm for}\;\; 
   X^+ <\alpha , X^- \geq 0, 
\end{equation}
and the spacetime is flat.
For this region the null line $X^-=0$  is part of ${\cal I}^+_L$. Similarly,
${\cal I}^-_R$ is found to be $X^- = \infty$.  Consideration of $X^+ >\beta$
fixes ${\cal I}^+_R$ to be at $X^+ = \infty$.

Next we examine the locus of the singularity:
\begin{eqnarray}
y(x) = 0 \Rightarrow & \kappa^2 X^+ X^-= 
                 \int^{X^+}_0 d\bar X{}^+
                    \int^{\bar X{}^+}_0
d\bar{\bar X}^+\left(f_{,+}(\bar{\bar X}^+)\right)^2 \\
  \Rightarrow X^- = &  {1\over \kappa^2 X^+}  
                      \int^{X^+}_0 d\bar X{}^+
                    \int^{\bar X{}^+}_0
d\bar{\bar X}^+\left(f_{,+}(\bar{\bar X}^+)\right)^2 \ .
\label{eq:singular}
\end{eqnarray}
The singularity intersects ${\cal I}^+_L$ at $(X^-=0, X^+= \alpha )$.
It can be checked that the normal $n_a= \partial_a y$
 to the curve corresponding to the singularity has norm in the auxillary 
metric given by 
\begin{eqnarray}
n^an_a &  = & \kappa^2 \left(
          \int^{X^+}_0 d\bar X{}^+
                    \int^{\bar X{}^+}_0
d\bar{\bar X}^+\left(f_{,+}(\bar{\bar X}^+)\right)^2
                     \;- \; X^+\int_{0}^{X^+}d\bar X{}^+ 
               \left(f_{,+}(\bar X{}^+)\right)^2 \right) \\
 & = &  -\kappa^2 \int_{0}^{X^+}d\bar X{}^+ \bar X{}^+ 
               \left(f_{,+}(\bar X{}^+)\right)^2 
\end{eqnarray}
This is clearly negative  for $X^+>\alpha $. Thus the singularity is
 spacelike.
 From (\ref{eq:singular}), the 
singularity `intersects' future right null infinity at 
\begin{equation}
X^-= {1\over \kappa^2} \int_{0}^{\infty}d X{}^+ 
               \left(f_{,+}( X{}^+)\right)^2 \  .
\label{eq:horizon}
\end{equation}
(\ref{eq:horizon})
 gives the position of the horizon for the black hole formed by collapse
of left moving matter. 

That there is a single spacelike curve solving (\ref{eq:singular}) can 
be seen from the following argument.\footnote{I thank J. Samuel for
 suggesting this argument.} Consider 
$y$ for fixed $X^-$ as a function of $X^+$. Let $X^+=X^+_{sing}>\alpha$ solve 
(\ref{eq:singular}). It can be checked that for $X^+>X^+_{sing}$, $y_{,+}<0$.
Thus, for a given $X^-$, $y=0$ occurs at a single value of $X^+$.

This completes the discussion of the physical spacetime.
As mentioned before, this solution admits an anlaytic extension to the 
whole $X^{\pm}$ plane. We now analyse this extension.

The full Minkowskian plane is divided into

\noindent (1)$X^{\pm}>0$: The physical spacetime lies within this range. 
It has an analytic extension `above' the singularity in which 
 $y<0$ and the metric acquires the signature +- instead of -+.

\vspace{3mm}
\noindent (2)$X^- >0 \ ,X^+ <0 $: (\ref{eq:y1l+}) gives 
$y=\kappa^2 X^+ X^-$. This describes a (complete) flat spacetime with 
$y<0$ (there is a `signature flip' for the analytically continued metric).

\vspace{3mm}
\noindent (3)$X^{\pm} <0$: $y=\kappa^2 X^+ X^-$ describes a complete flat 
spacetime with $y>0$.

\vspace{3mm}
\noindent (4)$ X^-<0 \ ,X^+ >0$:
Both terms on the right hand side of (\ref{eq:y1l+}) survive and both are
negative. So there is a `signature flip' in this region with 
$y<0$.

The structure in the full Minkowskian plane is shown, schematically, 
 in figure 1.

\section{Canonical description on the entire Minkowskian plane}

We describe the one sided collapse situation in classical 
canonical language. This is acheived by switching off degrees of freedom 
associated with the right moving matter fields, in a consistent manner,
in the description of \cite{jkm}.
Hence, we shall 
use the results and the  notation, and adapt the procedures, of 
\cite{jkm}. Rather than repeat the content of that paper here,
 we refer the reader to \cite{jkm}. Henceforth we shall assume familiarity
with that work. We shall also use results from \cite{karel} regarding the 
canonical transformation to the Heisenberg picture. Although that work
dealt with a spacetime topology $S^1 \times R$, 
the transformation to the Heisenberg picture
as well as other basic ideas such as the relation of canonical data with the 
spacetime solution of the Klein Gordon equation 
go through in the $R^2$ case which is of relevance here.

It would be straightforward, in what follows, to use the gauge 
fixing procedure
of \cite{mikovic}. Unfortunately, the gauge fixing conditions 
(\ref{eq:ggefix})
 {\em in conjunction with} the asymptotic conditions of \cite{jkm} result
in a foliation inappropriate for the entire Minkowskian plane. More precisely,
the foliation consists of boosted planes all passing through $X^+=X^-=0$ and
does not cover the timelike wedges $X^+X^- <0$.\footnote{ This does not 
necessarily rule out the existence of a {\em different} set of asymptotic 
conditions, which together with the same gauge fixing conditions 
(\ref{eq:ggefix}), gives a foliation which covers the entire Minkowskian 
plane.}
Such a foliation {\em does} cover the region $X^\pm >0$ and this why we use 
it in section IV.

In what follows, $x$ is a coordinate on the constant $t$ spatial slice
and the 1+1 Hamiltonian decomposition is in the context of a foliation 
of spacetime by such slices.
We use notation such that for a given field $g(x,t)$, 
${\partial g \over \partial x}$ 
is denoted by $g^\prime$ and 
${\partial g \over \partial t}$ is denoted by $\dot g$.

\subsection{Classical theory}

In \cite{jkm} the CGHS model is mapped to parametrized free field theory on 
a flat 2d spacetime. The  transformation from (\ref{eq:S}) 
(after parametrization
 at infinities) is made to a description 
in terms of embedding variables and the (canonical form of the)
action in these variables is
\begin{eqnarray}
S[X^\pm,\Pi_\pm,f,\pi_f,{\bar N},N^1;p,m_R)\;\;\;\;\;\;\;\;\;\;\;\;\;\;\;\;
\;\;\;\;\;\;\;\;\;\;\;\;\;\;\;\;
\nonumber\\
=\int dt\int_{-\infty}^{\infty}dx\ \left(\Pi_+\dot X{}^+ +
\Pi_-\dot X{}^-+\pi_f\dot f-{\bar N}{\bar H} - N^1 H_1\right)
 +\int dt\ p\dot m_R\ . 
\label{eq:Sfinal}
\end{eqnarray}
Here $X^{\pm}$ are the embedding variables (they correspond to the light 
cone coordinates we have been using to describe the solution in earlier 
sections), $\Pi_\pm$ are their conjuagte momenta, $f$ is the scalar field and
$\pi_f$ it's conjugate momentum, ${\bar N}$ and $N^1$ are the rescaled
lapse and the shift and 
${\bar H}$ and $H_1$ are the rescaled super-Hamiltonian and 
supermomentum constraints which take the form of the constraints for a 
parametrized massless scalar field on a 2-dimensional Minkowski spacetime. 
It is convenient to deal with the Virasoro combinations
\begin{equation}
H^\pm:={1\over 2}(\bar H\pm H_1)
=\pm\Pi_\pm X^\pm{}' +{1\over 4}(\pi_f\pm f')^2\approx 0\ .
\label{eq:vc}
\end{equation}
$m_R$ is the right mass of the spacetime and $p$, its conjugate momentum,
has the interpretation of the difference between the parametrization
time and the  proper time at right spatial infinity with the left 
parametrization time chosen to agree with the left proper time.
It is useful to recall from \cite{jkm} that 
\begin{eqnarray}
y(x)&=&{\kappa}^2 X^+(x)X^-(x) 
\nonumber\\
&&-\int^{x}_{\infty}d\bar x\ X^-{}'(\bar x)
\int^{\bar x}_{\infty}d\bar{\bar x}\ \Pi_-(\bar{\bar x})
+\int^{x}_{-\infty}d\bar x\ X^+{}'(\bar x)
\int^{\bar x}_{-\infty}d\bar{\bar x}\ \Pi_+(\bar{\bar x})
\nonumber\\
&&+\int^{\infty}_{-\infty}dx\ X^+(x)\Pi_+(x) +{m_R\over\kappa}\ .
\label{eq:y(X)}
\end{eqnarray}
Note that the right mass is related to the left mass by 
\begin{equation}
{m_L\over\kappa}={m_R\over\kappa}
+\int_{-\infty}^\infty dx\ X^+(x)\Pi_+(x)
-\int_{-\infty}^\infty dx\ X^-(x)\Pi_-(x)\ .
\label{eq:mL}
\end{equation}
That the right mass appears in (\ref{eq:Sfinal})
rather than the left mass $m_L$ is a matter of choice. In \cite{jkm} if 
the authors 
had chosen to synchronise the right parametrization clock with the right 
proper time, the last term in (\ref{eq:Sfinal}) would be 
$\int dt{\bar p}\dot m_L$
where $\bar p$ denotes the difference between the left parametrization and 
the 
left proper time.
So an action equivalent to (\ref{eq:Sfinal}) is 
\begin{eqnarray}
S[X^\pm,\Pi_\pm,f,\pi_f,{\bar N},N^1;{\bar p},m_L)\;\;\;\;\;\;\;\;\;\;\;\; 
\;\;\;\;\;\;\;\;\;\;\;\; 
\nonumber\\
=\int dt\int_{-\infty}^{\infty}dx\ \left(\Pi_+\dot X{}^+ +
\Pi_-\dot X{}^-+\pi_f\dot f-{\bar N}{\bar H} - N^1 H_1\right)
+\int dt{\bar p}\dot m_L\ .
\label{eq:Sfinall}
\end{eqnarray}

Note that $m_L, {\bar p}$ are constants of motion. To make contact with the 
solution in Section II.3, we 
first freeze the left mass to zero and simultaneously
 put  ${\bar p}=0$. The reduced action, without this pair,  
\begin{eqnarray}
S[X^\pm,\Pi_\pm,f,\pi_f,{\bar N},N^1;{\bar p}=0,m_L=0)\;\;\;\;\;\;
\;\;\;\;\;\;\;\;\;\;
\nonumber\\
=\int dt\int_{-\infty}^{\infty}dx\ \left(\Pi_+\dot X{}^+ +
\Pi_-\dot X{}^-+\pi_f\dot f-{\bar N}{\bar H} - N^1 H_1\right) 
\label{eq:Sfinalred}
\end{eqnarray}
reproduces the correct equations of motion. 
Alternatively, if one is not familiar with the parametrization at infinities
procedure, it can be checked from (\ref{eq:Sfinal}) that $m_R, p$ are 
constants of motion. We can, therefore, consistently freeze this degree
of freedom by setting $m_R, p$ equal to {\em constants} of motion 
and then simply use the reduced action (\ref{eq:Sfinalred}). Since 
\begin{equation}
{m\over \kappa}:=-\int_{-\infty}^\infty dx\ X^+(x)\Pi_+(x)
+\int_{-\infty}^\infty dx\ X^-(x)\Pi_-(x) 
\label{eq:m}
\end{equation}
commutes with the constraints, 
it is a constant of a motion and we can consistently set $m_R =m$ and $p=0$.

Next, in order to have a description of the one sided collapse situation,
we must set $f^-=0$. This is done in the canonical treatment as follows.
 Through a Hamilton-Jacobi type of transformation we pass from
the description in (\ref{eq:Sfinalred}) to the Heisenberg picture 
\cite{jkm,karel}. The new variables are the Fourier modes (they can be 
interpreted as determining the matter field and momentum on an initial slice)
$a_\pm (k)$ ($k >0 $ ), 
their  complex conjugates $a^*_\pm (k)$, the embedding
variables $X^{\pm}(x)$ (these are unchanged) and the new embedding 
momenta ${\bar \Pi}_{\pm}$. The Fourier modes and the new embedding momenta 
are given by  
\begin{eqnarray}
a_\pm (k) & = & {i\over 2\sqrt{\pi k}}\int_{-\infty}^{\infty} 
                   (\pi_f \pm f^\prime )e^{ikX^+(x)}\ ,  
\label{eq:ak}
\\
{\bar \Pi}_{\pm} & = & {H^{\pm}\over X^{\pm \prime}} \ .
\label{eq:Pi}
\end{eqnarray}
Thus, the vanishing of the constraints is equivalent to the vanishing of the 
new embedding momenta. The only nontrivial Poisson brackets for the new 
variables are 
\begin{equation}
\{a_\pm (k), a^*_\pm (l)\} =-i \delta (k,l)\;\;\;
\{X^\pm (x), {\bar \Pi}_{\pm}(y)\}= \delta (x,y).
\end{equation}
To summarize, the scalar field and momenta are replaced by their Fourier
modes which can be thought of as coordinatizing their values on an initial
slice given by \\
\noindent $X^+(x)-X^-(x)=0$. The embedding coordinates are unchanged and
the new embedding momenta are essentially the old constraints.

Setting $f^-=0$ is equivalent in the canonical language to demanding
$\pi_f - f^{\prime}=0$ \cite{karel}. From (\ref{eq:ak}), this is 
achieved by setting  $a_-(k)=a^*_-(k)=0$ and this can be done 
consistently, since the `+' and `-' modes are not dynamically coupled. 
So the final variables for the theory are 
$a_+(k), a^*_+(k), X^{\pm}(x)$ and ${\bar \Pi}_{\pm}(x)$. The latter
vanish on the constraint surface. The connection to the 
variables $X^{\pm},\Pi_{\pm}, f, \pi_f$ 
\cite{jkm} is through (\ref{eq:ak}) and (\ref{eq:Pi}). The 
$X^{\pm},\Pi_{\pm}$  variables are themselves related to the 
geometric variables of interest (the dilaton and its canonically conjugate
momentum, the induced metric on the spatial slice and its conjugate momentum)
in \cite{jkm}. 

Since we are dealing with $-\infty <X^\pm < \infty $ \cite{jkm}, this
analysis (and the next two sections on quantum theory) 
pertains to the analytically
extended one sided collapse solution.

In this paper we shall examine only the dilaton field 
(\ref{eq:y(X)}). As mentioned in \cite{jkm}, solving the constraints, 
$H^{\pm}$, 
expresses $\Pi_{\pm}$ in terms of $X^{\pm}$ and the scalar field and its 
momentum. Substituting this in (\ref{eq:y(X)}) and using 
$\pi_f + f'= 2X^{+\prime}f_{,+}$ 
from \cite{karel}, one is led back to the spacetime
solution
\begin{equation}
y(X)=\kappa^2 X^+ X^- -\int^{X^+}_{-\infty} d\bar X{}^+
                    \int^{\bar X{}^+}_{-\infty}
d\bar{\bar X}^+\left(f_{,+}(\bar{\bar X}^+)\right)^2 
\label{eq:y1l+inf}
\end{equation}

For the next section it is useful to examine the 
large $X^+$ behaviour of (\ref{eq:y1l+inf}). 
For $X^+$ large enough that it is outside the support of the matter
\begin{equation}
y = \kappa^2 X^+ X^-  - X^+H + {m_R \over \kappa}
\label{eq:yinfty}
\end{equation}
with 
\begin{equation}
H= \int^{\infty}_{-\infty} d\bar X{}^+
                                \left(f_{,+}({\bar X}^+)\right)^2 
\label{eq:H}
\end{equation}
and $m_R$ given by the right hand side of (\ref{eq:m}).

\subsection{Quantum theory}
The passage to quantum theory is straightforward. From \cite{jkm},
 the operators 
${\hat X}^{\pm}$ are represented by multiplication, 
${\hat \Pi}_{\pm}=-i\hbar{\delta \over \delta X^\pm}$ and 
${\hat a}_+(k),{\hat a}^{\dagger}_+(k)$ by representation on a Fock space.
Note that $\sqrt{-k}{\hat{\bf a}}(-k)$, ($k>0$),
 in \cite{jkm} corresponds, here, to
${\hat a}_+(k)$ and that the commutator
\begin{equation}
[{\hat a}_+(k), {\hat a}^{\dagger}_+(l)] = \hbar \delta(k,l) \ .
\end{equation}
The imposition of the quantum version of the classical Heisenberg picture
constraints leads us to the quantum Heisenberg picture, wherein states lie
in the standard embedding independent Fock space. Note that the Fock space
here is spanned by the restriction of the Fock basis of \cite{jkm} to  
negative momenta because we have frozen the right moving modes.

In the next section, we show existence of large quantum gravity effects
at large $X^+$. This involves calculation of fluctuations of operators, 
$\hat Q$, 
of the form
\begin{equation}
{\hat Q}= \int_{-\infty}^{\infty}dX^+ Q(X^+) 
 :\left({\hat f}_{,+}(X^+)\right)^2:
\label{eq:Q}
\end{equation}
where $Q(X^+)$ is a c-number function,  `::' refers to normal ordering  and 
\begin{equation}
{\hat f}(X^+) = {1\over 2\sqrt{\pi}}\int_0^{\infty}{dk\over\sqrt{k}}
                 \left({\hat a}_+(k)e^{-ikX^+} +
{\hat a}^{\dagger}_+(k)e^{ikX^+} \right) \ .
\label{eq:fquantum}
\end{equation}
Consider the coherent state
\begin{equation}
|\psi_c> = e^{-{1\over 2\hbar}\int_{0}^{\infty}|c_+(k)|^2 dk}
                         \exp\left(\int_0^{\infty}{dk\over\hbar}
            c_+(k){\hat a}^{\dagger}_+(k)\right) |0> 
\end{equation}
where $|0>$ is the Fock vacuum 
 and $c_+(k)$ are the (c-number) modes
of the classical field $f_c(X^+)$,
\begin{equation}
 f_c(X^+) = {1\over 2\sqrt{\pi}}\int_0^{\infty}{dk\over\sqrt{k}}
                 \left(c_+(k)e^{-ikX^+} +
 c^{*}_+(k)e^{ikX^+} \right) \ .
\label{eq:fclassical}
\end{equation}
The mean value $\bar Q$ of the operator $\hat Q$ in this coherent state is
given by
\begin{equation}
<\psi_c| {\hat Q}|\psi_c> 
  = \int_{-\infty}^{\infty}dX^+ Q(X^+)
                                 \left(f_{c,+}(X^+)\right)^2
\label{eq:Qbar}
\end{equation}
as expected.

The (square) of the fluctuation in $\hat Q$ is given by 
\begin{eqnarray}
(\Delta Q)^2 & = & <\psi_c| {\hat Q}^2|\psi_c> - {\bar Q}^2 \nonumber\\
         & = & { {\hbar}^2\over 8\pi}\int_{0}^{\infty}dk k^3|Q(k)|^2
 +{\hbar\over 4}\int_{0}^{\infty} dk k |Q_f(k)|^2 
\label{eq:DeltaQ} 
\end{eqnarray}
where $Q(k)$ is the Fourier transform of $Q(X^+)$ and $Q_f(k)$ is the 
Fourier transform of the function $Q_f(X^+) := Q(X^+)f_{c,+}(X^+)$. The 
Fourier transform of a function $g(X^+)$ is 
\begin{equation}
g(k) = {1\over \sqrt{\pi}}\int_{-\infty}^{\infty}dX^+ e^{ikX^+}g(X^+) \ .
\end{equation}
Note that by virtue of its being independent of $f_c(X^+)$ 
the $\hbar^2$ term in (\ref{eq:DeltaQ}) is the vacuum fluctuation of $\hat Q$.

\subsection{Large Quantum Gravity effects}
We examine the fluctuations of the dilaton field, $y$, which plays the 
role of a conformal factor for the physical metric and hence encodes all 
the nontrivial metrical behaviour.
The expression for $y$ simplifies  at large $X^+$ and we shall 
calculate the fluctuations of $\hat y$ in this limit.
$y(X)$ is turned into the operator ${\hat y} (X)$
by substituting the appropriate embedding dependent Heisenberg 
field operators (\ref{eq:fquantum}) in (\ref{eq:y1l+inf}). Similarly
$H$ and $m_R$ are turned into the operators $\hat H$ and $\hat m_R$.
Note that $y(x)$
is not a Dirac observable. However it can be turned into one using the 
`evolving constants of motion' interpretation (see \cite{evcons} and 
references therein).

Straightforward calculations result in the following expression for the 
ratio of the fluctuation in $\hat y$ to its mean value, at large $X^+$
\begin{equation}
\left({\Delta y \over {\bar y}}\right)^2 =
     {  
       (\Delta H)^2 + \left({\Delta m_R\over\kappa X^+}\right)^2-
                   {1\over\ X^+ \kappa}\left( \bar{[{\hat H},{\hat m_R}]_+}
                                               -2{\bar H}{\bar m_R}    
                                         \right)
                          \over         
       (\kappa^2 X^{-})^2 \left(1- {{\bar H}\over \kappa^2 X^-}+ 
                                       {{\bar m_R}\over \kappa^3 X^+ X^-}
                                       \right)^2  
        }    .
\label{eq:Deltay}
\end{equation}
Here 
\begin{equation}
\bar{[{\hat H},{\hat m_R}]_+}= 
                <\psi_c| {\hat H}{\hat m_R}+ {\hat m_R}{\hat H}|\psi_c>  .
\label{eq:anticom}
\end{equation}
To make contact with the classical solution of section II, we choose the 
coherent state to be such that $f_c(X^+)$ is of compact support. Further 
\begin{equation}
f_c(X^+) =0 \; {\rm for} \; X^+ \leq 0 .
\end{equation}
$\hat H$ corresponds to choosing $Q(X^+)=:H(X^+)= 1$ in (\ref{eq:Q})
and ${{\hat m_R}\over \kappa}$ to \\
\noindent $Q(X^+)=:{ m_R(X^+)\over \kappa}=X^+$.
From (\ref{eq:Qbar}), and the fact that $f_c(X^+)$ is of compact support,
it is easy to see that ${\bar H}, {\bar m_R}$
are finite.

Using (\ref{eq:DeltaQ}) it can be seen (in obvious notation) that 
\begin{equation}
(\Delta H)^2  =
     {\hbar\over 4}\int_{0}^{\infty} dk k |H_f(k)|^2 
              =
     \hbar\int_{0}^{\infty} dk k^2 |c_+(k)|^2 .
\label{eq:DeltaH}
\end{equation}
Since $f_c$ is of compact support, 
its Fourier modes decrease rapidly at
infinity and have sufficiently good infrared behaviour that the 
integral above  is both ultraviolet as well as infrared finite. 
This shows that the fluctuation in $\hat H$ is finite.

The fluctuation in $\hat m_R$ is 
\begin{equation}
(\Delta m_R)^2 
          =  { {\hbar}^2\over 8\pi}\int_{0}^{\infty}dk k^3|m_R(k)|^2
 +{\hbar\over 4}\int_{0}^{\infty} dk k |m_{Rf}(k)|^2 . 
\label{eq:Deltam}
\end{equation}
Again, the fact that $f_c$ is of compact support renders the second term
on the right hand side of (\ref{eq:Deltam}) UV and IR finite.
We now argue that the first term corresponding to the vacuum fluctuation,
\begin{equation}
(\Delta_0 m_R)^2 
          :=  { {\hbar}^2\over 8\pi}\int_{0}^{\infty}dk k^3|m_R(k)|^2
\label{eq:Deltam0}
\end{equation}
is finite. Since $m_R(X^+)= \kappa X^+$, its Fourier transform $m_R(k)$
is ill defined. We calculate, instead, the vacuum fluctuation of the 
regulated operator ${\hat m}_R^{(D)}$ defined by setting 
\begin{equation}
m_R^{(D)}(X^+)= \kappa X^+ e^{{-(X^+)^2 \over D^2}}  . 
\end{equation}
We shall take the $D\rightarrow\infty$ limit at the end of the calculation
to obtain the vacuum fluctuation of 
${\hat m}_{R}={\hat m}_{R}^{(\infty )}$.

Now $m_{R}^{(D)}(X^+)$ is a function of sufficiently rapid decrease
at infinity  that  
$(\Delta_0 m_{R}^{(D)})^2 $ exists. It is evaluated to be
\begin{equation}
(\Delta_0 m_{R}^{(D)})^2 = {\hbar^2 \kappa^2 \over 6} \ ,
\end{equation}
which is finite and independent of $D$. Thus the $D\rightarrow \infty$
limit can be taken and we have 
\begin{equation}
(\Delta_0 m_{R})^2 = {\hbar^2 \kappa^2 \over 6} \ .
\end{equation}
Finally, a straightforward calculation shows 
(\ref{eq:anticom}) also to be finite.

\noindent We evaluate (\ref{eq:Deltay}) for 2 cases:

\noindent {\bf Case I} 
Near right spatial infinity: 

Here $X^\pm\rightarrow\infty$.
So
\begin{equation}
\left({\Delta y \over {\bar y}}\right)^2 \rightarrow 0
\end{equation}
as $O({1\over (X^-)^2})$. Thus, unlike the cylindrical wave case \cite{abhay},
there are no large quantum fluctuations of the metric near spatial infinity.
This is because the leading order behaviour of the metric is dictated by
$\kappa^2 X^+ X^-$ which is a state independent c-number function, unlike 
in the cylindrical wave case.

\vspace{3mm}

\noindent {\bf Case II}$\; X^- = {{\bar H}\over \kappa^2} -
                      {{\bar m_R}\over \kappa^3 X^+} +d$ and $X^+$ large:

Here $d$ is a real parameter. It can be checked that $d$ 
measures the  distance  
in $X^-$ from the singularity which occurs at $d=0$ 
(see (\ref{eq:singular})).
It is easy to see that 
\begin{equation}
\left({\Delta y \over {\bar y}}\right)^2  = {(\Delta H)^2\over 
                                  \kappa^4 d^2} + O({1\over (X^+)^2})
\label{eq:flucsing}
\end{equation}
This expression makes {\em no} assumptions on the size of $d$. Using 
(\ref{eq:DeltaH}) in (\ref{eq:flucsing}) and  reinstating explicitly
the factors of $G$ \footnote{Units are discussed in \cite{jkm}}
(and keeping $c=1$), we find that upto leading order in $X^+$
\begin{eqnarray}
\left({\Delta y \over {\bar y}}\right)^2 &  = & 
        {\hbar G^2\over \kappa^4d^2}\int_{0}^{\infty} dl l^2 |c_+(l)|^2 \\
   &  =  & {\hbar G\over \kappa^2 d^2}
          \int_{0}^{\infty} d{l\over\kappa} ({l\over\kappa})^2 
                           (G\kappa |c_+(l)|^2) .
\label{eq:flucsingG}
\end{eqnarray}
Note that in $c=1$ units, $[G]= M^{-1}L^{-1}$, $[\kappa ]= L^{-1}$,
$[c_+(l)]= M^{{1\over 2}}L$ and $[\hbar ]=ML$. 
Thus $\hbar G$ is the dimensionless `Planck number', and 
$\kappa d,\;{l\over\kappa}$ and $G\kappa |c_+(l)|^2$ are all dimensionless.
From (\ref{eq:flucsingG}), there are large fluctuations in ${\hat y}$ when 
\begin{equation}
\hbar G\;>>\; {\kappa^2 d^2 
             \over \int_{0}^{\infty} d{l\over\kappa} ({l\over\kappa})^2 
                          (G\kappa |c_+(l)|^2)}
\label{eq:ineq}
\end{equation}
This does not require $d$ to be small! Large fluctuations can occur
even if $\kappa d >>1$, provided the integral in (\ref{eq:ineq}) is large 
enough. Two cases when this 
is possible is if there are large enough number of low frequency
scalar field excitations or if there is a high frequency `blip' in the 
scalar field. This is very similar to what happens in \cite{abhay}.
Note that on a classical solution with mass $m_R$, the classical
scalar curvature at a `distance' $d$ from the singularity as a function of 
$X^+$, is at large enough $X^+$, 
\begin{equation}
R= {m_R\over \kappa  X^+ d}
\end{equation}
and vanishes at $X^+=\infty$.

The horizon  is located (approximately) at 
$X^-_{H} := {{\bar H}\over \kappa^2}$ 
(\ref{eq:horizon}). Therefore, if $d<0$, the region under consideration lies 
within $X^-_{H}$; if $d>0$ and $X^+$ is large enough, the region lies 
outside $X^-_{H}$. Thus, for states satisfying (\ref{eq:ineq}), large 
quantum fluctuations in the metric occur both within and outside the 
`mean ' location of the horizon. But from (\ref{eq:horizon})
this location itself fluctuates by 
${\Delta H\over \kappa^2}$ ! Thus, if the Planck number is much less
 than one, the above calculation does not 
show existence of 
large quantum fluctuations {\em outside} the {\em fluctuating} horizon.
\footnote{I thank Sukanta Bose for comments regarding this point.}
\section{Canonical description on the $X^{\pm}>0$ sector of the Minkowskian 
      plane}
Section III is applicable to the analytic extension of 
the one sided collapse situation to the full Minkowskian plane. In this 
section
we attempt to deal {\em only} with the physical spacetime and not with its
analytic extension. 
We modify the analysis of \cite{jkm}, pertinent to the entire Minkowskian
plane $ -\infty < X^{\pm} < \infty$ in order to treat the case  when 
$X^{\pm}>0$. This involves a modification of the asymptotics at left spatial 
infinity. Now, $x$ is restricted to be positive and $x=0$ labels left 
spatial infinity. As mentioned in section III, 
the simplest route to quantum theory is through gauge fixing
\cite{mikovic} the description in terms of the original geometric variables
rather than by transforming to embedding variables.
\subsection{Classical theory}
As in the previous section we assume familiarity with \cite{jkm}.
The canonical form of the action in the original geometric variables is 
\begin{eqnarray}
\lefteqn{S[y,\pi_y,\sigma,p_\sigma,f,\pi_f,{\bar N},N^1]}\nonumber\\
&=&\int dt\int_{-\infty}^\infty dx\ \left(\pi_y\dot y+p_\sigma\dot\sigma
+\pi_f\dot f-{\bar N}{\bar H} - N^1H_1\right)
\label{eq:Sy}
\end{eqnarray}
with 
\begin{equation}
\bar H=-\pi_y\sigma p_\sigma + y''-\sigma^{-1}\sigma'y'
- 2 \kappa^2\sigma^2 +{1\over 2}(\pi_f^2+f'{}^2)
\end{equation}
and 
\begin{equation}
H_1=\pi_y y' - \sigma p_\sigma' + \pi_f f'  \ .
\end{equation}
Here $\pi_y$ is the momentum conjugate to the dilaton, $\sigma$ is the 
spatial metric (induced from the auxilliary spacetime metric) and $p_{\sigma}$
is its conjugate momentum.

The asymptotic conditions at right spatial infinity (which corresponds to 
$x=\infty$) are unchanged from \cite{jkm}.
Left spatial infinity is labelled by $x=0$. We require, as $x\rightarrow 0$
\begin{eqnarray}
y = \kappa^2 x^2 + O(x^3) & \sigma = 1 +O(x^2) \\
\label{eq:yflat}
\pi_y = O(x)  & p_\sigma = O(x^2) \\
{\bar N} =\alpha_L x + O(x^3) & N^1 =O(x^3) \ ,
\end{eqnarray}
where $\alpha_L$ is a real parameter.

(\ref{eq:Sy}) is augmented with surface terms to render it functionally
differentiable. The result is 
\begin{eqnarray}
\lefteqn{S[y,\pi_y,\sigma,p_\sigma,f,\pi_f,{\bar N},N^1]}\nonumber\\
&=&\int dt\int_{0}^\infty dx\ \left(\pi_y\dot y+p_\sigma\dot\sigma
+\pi_f\dot f-{\bar N}{\bar H} - N^1H_1\right)\nonumber\\
&&\quad\quad\quad +\int dt\ \left(-\alpha_R{m_R\over\kappa}\right)\ .
\label{eq:Sbdry}
\end{eqnarray}
Here $\alpha_R$ is related to the asymptotic behaviour of the lapse at 
right spatial infinity\cite{jkm}. 
It can be checked that with $f,\pi_f$ of compact support, all the 
asymptotic  conditions are preserved under evolution. Note that 
(60) automatically ensures that $m_L=0$. 

To make contact with the 1 sided collapse solution the right moving modes
must be set to zero. We do this as follows. Note that 
upto total time derivatives
\begin{equation}
2\int_{0}^{\infty} \pi_f {\dot f} 
  = \int_{0}^{\infty}dx \left(\int_{0}^x \pi_-({\bar x})d{\bar x}\right) 
                                                   {\dot {\pi_-} (x)}
-\int_{0}^{\infty}dx \left(\int_{0}^x \pi_+({\bar x})d{\bar x}\right) 
                                                   {\dot {\pi_+} (x)}
\label{eq:fsympl}
\end{equation}
where 
\begin{equation}
\pi_{\pm}:= \pi_f \pm f' \ .
\end{equation}
Thus, we can replace $f, \pi_f$ by $\pi_\pm$, with the new Poisson brackets
being 
\begin{equation}
\{\pi_{\pm}(x), \pi_{\pm}(y)\} =\pm 2 {d\delta (x,y)\over dx}
\;\;\;\;\;
\{\pi_{+}(x),\pi_{-}(y)\} =0 \ .
\label{eq:pipb}
\end{equation}
Since $\pi_+$ and $\pi_-$ do not couple dynamically, 
we can consistently freeze $\pi_-=0$.
This corresponds to setting $f^-=0$. 

Now, with a view towards quantization we introduce the gauge fixing conditions
\cite{mikovic}
\begin{equation}
\pi_y =0 \;\;\; \sigma=1 \ .
\label{eq:ggefix}
\end{equation}
Using (\ref{eq:ggefix}) in the constraints, the general solution for $y$ and
$p_\sigma$, in terms of $\pi_+$, consistent with the asymptotic conditions 
at left and right spatial infinity is:
\begin{eqnarray}
y & = & \kappa^2 x^2 - \int_0^x d{\bar x}\int_0^{\bar x}
                                d{\bar {\bar x}}{\pi_+^2({\bar {\bar x}})
                                                 \over 4} \\
\label{eq:yggefix}
p_\sigma & = & \int_0^x d{\bar x} {\pi_+^2({\bar x})\over 4} \ .
\label{eq:psigggefix}
\end{eqnarray}
Requiring preservation of (\ref{eq:ggefix})
under evolution  along with consistency with the asymptotic conditions
fixes
\begin{equation}
{\bar N}= \alpha x \;\;\; N^1 =0 
\end{equation}
where $\alpha$ is a real parameter.
From \cite{jkm} and (68)
\begin{equation}
{m_R\over\kappa} = \int_0^\infty dx x {\pi^2_+(x)\over 4} \ .
\end{equation}
Substituting this in (\ref{eq:Sbdry}) and using (\ref{eq:fsympl}), we get 
\begin{equation}
S[\pi_+(x)] = 
-\int dt\left(\int_{0}^{\infty}dx \left(\int_{0}^x {\pi_+({\bar x})\over 2}
              d{\bar x}\right) {\dot {\pi_+} (x)}
              - \int_0^\infty dx x {\pi^2_+(x)\over 4}\right) \ .            
\end{equation}
In the above equation put
\begin{equation}
\kappa r := \ln (\kappa x) \;\;\;\; {\bar \pi}_+= e^{\kappa r}\pi_+
\end{equation}
to get 
\begin{equation}
S[{\bar \pi}_+(x)] = 
-\int dt\left(\int_{-\infty}^{\infty}dr 
\left(\int_{-\infty}^r {{\bar \pi}_+({\bar r})\over 2}
              d{\bar r}\right) {\dot {{\bar \pi}_+} (r)}
           - \int_{-\infty}^\infty dr  {{\bar \pi}^2_+(r)\over 4}\right) \ .
\end{equation}
Note that the last term ($={m_R\over \kappa}$)
simplifies.
The equations of motion are 
\begin{equation}
{\dot {{\bar \pi}_+} (r,t)}
=  \{{\bar \pi}_+ (r,t), {m_R\over \kappa}\}
= {\partial{\bar \pi}_+ (r,t)\over \partial r}
\end{equation}
The appropriate mode expansion which solves this is
\begin{equation}
{\bar \pi}_+(r,t) = {1\over \sqrt{\pi}}\int_0^{\infty}dk\sqrt{k}
                 \left(-i{\bar a}_+(k)e^{-ikr^+} +
i{\bar a}^{*}_+(k)e^{ikr^+} \right)
\label{eq:pi+mode}
\end{equation}
where $r^+:= r+t$. From (\ref{eq:pipb}) and (\ref{eq:pi+mode}), the 
only nontrivial Poisson brackets between the mode coefficients are
\begin{equation}
\{ {\bar a}_+(k), {\bar a}^{*}_+(l)\} = (-i)\delta (k,l)
\label{eq:barapb}
\end{equation}
From \cite{jkm}, one can understand the slicing 
of the spacetime corresponding to the gauge fixing conditions
(\ref{eq:ggefix}) we have used. In particular, on a solution, 
one can  see that 
$(r,t)$ is related to $X^\pm$ by 
\begin{equation}
\kappa X^\pm = e^{\kappa (r\pm t)}
\end{equation}
and that the physical metric in these coordinates is manifestly 
asymptotically 
flat at the spatial infinities. 

The large $X^+$ behaviour of $y$ is again given by (\ref{eq:yinfty}). 
Now $y,H$
and ${m_R\over \kappa}$ evaluated in the new coordinates take the form
\begin{equation}
y = \kappa^2 e^{2\kappa r}  - e^{\kappa r^+}H + {m_R \over \kappa} \ ,
\label{eq:yinftyr}
\end{equation}
\begin{equation}
H
= \int^{\infty}_{0} d X{}^+
                                {( \pi_{+}( X^+))^2\over 4} 
= \int^{\infty}_{-\infty} d r{}^+ e^{-\kappa  r^+}
                                {( {\bar \pi}_{+}(r^+))^2\over 4} 
\label{eq:Hr}
\end{equation}
and 
\begin{equation}
{m_R\over \kappa} = = \int^{\infty}_{0} dX{}^+ X^+
                                ( \pi_{+}( X^+))^2 
= \int_{-\infty}^\infty dr^+ {{\bar \pi}^2_+(r^+)\over 4} \ .       
\label{eq:mRr}
\end{equation}
In the $X^{\pm}$ coordinates $H$ (\ref{eq:H})
took the form of the conventional Hamiltonian for free field theory on 
 the entire 
$X^{\pm}$ plane, but ${m_R\over \kappa}$ was more complicated.
In the $(r,t)$ coordinates, ${m_R\over \kappa}$ takes the form of the 
conventional Hamiltonian for free field  theory on the entire $(r,t)$
plane (this is just the $X^{\pm}>0$ part of the entire $X^{\pm}$ plane),
but $H$ is complicated.

\subsection{Quantum theory}
The mode operators 
${\hat {\bar a}}_+(k),{\hat {\bar a}}^{\dagger}_+(k)$
are represented in the standard way on the Fock space with 
vacuum $|{\bar 0}>$. 
They have the standard commutation relations 
\begin{equation}
[{\hat {\bar a}}_+(k),{\hat {\bar a}}^{\dagger}_+(l)] =\hbar \delta (k,l)\ .
\end{equation}

Following the pattern of section III.2 and III.3
we attempt to calculate the fluctuations of $\hat y$ in the large $X^+$ 
region, in the  coherent state 
\begin{equation}
|\psi_c> = e^{-{1\over 2\hbar}\int_{0}^{\infty}|{\bar c}_+(k)|^2 dk}
                         \exp\left(\int_0^{\infty}{dk\over\hbar}
            {\bar c}_+(k){\hat {\bar a}}^{\dagger}_+(k)\right) |{\bar 0}> 
\end{equation}
corresponding to the classical field
\begin{equation}
 {\bar \pi}_{+c}(r^+) = {1\over \sqrt{\pi}}\int_0^{\infty}dk\sqrt{k}
                 \left(-i{\bar c}_+(k)e^{-ikr^+} +
 i{\bar c}^{*}_+(k)e^{ikr^+} \right) \ .
\label{eq:piclassical}
\end{equation}
which is of compact support in $r^+$.

Formally, (\ref{eq:Deltay}) again expresses the fluctuations in ${\hat y}$.
However, now the crucial operator is $\hat H$.
It is obtained from the corresponding classical expression (\ref{eq:Hr}) 
in the obvious way. Similar calculations to those in section 3.2
give
\begin{eqnarray}
(\Delta H)^2 
         & = & { {\hbar}^2\over 8\pi}\int_{0}^{\infty}dk k^3|H(k)|^2
 +{\hbar\over 4}\int_{0}^{\infty} dk k |H_{\bar \pi}(k)|^2 
\label{eq:DeltaHr} 
\end{eqnarray}
where $H(k)$ is the Fourier transform of 
\begin{equation}
H(r^+):=e^{-\kappa r^+}
\end{equation}
 and 
$H_{\bar \pi}(k)$ is the 
Fourier transform of the function 
\begin{equation}
H_{\bar \pi}(r^+) := H(r^+){{\bar \pi}_{+c}(r^+)\over 4} \ .
\end{equation}
The Fourier transform of a function $g(r^+)$ is 
\begin{equation}
g(k) = {1\over \sqrt{\pi}}\int_{-\infty}^{\infty}dr^+ e^{ikr^+}g(r^+) \ .
\end{equation}
Since $e^{-\kappa r^+}$ is not a function of rapid decrease
in the $r^+$ variable, the first integral in 
(\ref{eq:DeltaHr}) is ill defined. It can be regulated by introducing
the regulator $e^{-{(r^+)^2\over D^2}}$. The relevant regulated integral
diverges in the limit $D\rightarrow \infty$ as ${e^{2D^2}\over D^2}$. 
Thus the vacuum fluctuations of $\hat H$  diverge.
Therefore,  in this representation we cannot proceed further with the analysis
and $\hat y$, as it stands, 
 cannot be given meaning as an operator on the Fock space.

\section{Discussion}

One way of describing the spacetime geometries which arise in the CGHS model
is as follows. Consider the Minkowskian plane with the flat auxilliary metric 
(\ref{eq:flatmetric}), on which a scalar field propagates in accordance with 
the flat space wave equation. The spacetime metric is conformal to the 
auxilliary flat one. The conformal factor $y$ is determined by the matter 
distribution through (\ref{eq:y1}) and is required to be positive. 
The field equations continue to make sense for $y\leq 0$. If one removes the 
restriction of positivity of $y$, then the following picture emerges. The 
Minkowskian plane is divided into spacetimes each of which have $y>0$ or 
$y<0$. The former have the signature $-+$ and the latter $+-$. As far as we 
know, typically, $y=0$ labels singularities or boundaries at infinity for 
these
spacetimes and some of these singularities may be past singularities.

This is the classical picture which corresponds to the quantum theory in 
\cite{jkm}. Among all these classical solutions there are solutions which 
describe black holes 
formed from matter collapse. It may be that the entire solution space and the 
associated quantum theory \cite{jkm,mikovic,jackiw} is required, in order to
understand issues arising in black hole formation. In particular it may be 
that
an understanding of Hawking radiation from a nonperturbative quantum theoretic
viewpoint requires a treatment as in \cite{mikovic}.

However, in this paper we have adopted the viewpoint that only the solutions 
which describe the physically interesting situation of black hole formation
through matter collapse are to be taken as the basis for passage to quantum
 theory. We have shown that these solutions have only left or right moving 
matter. We concentrated on the solution with left moving matter which 
described a collapsing black hole spacetime 
in the $X^{\pm} >0$ part of the plane. This solution admitted analytic 
extension to the full Minkowskian plane and we showed existence of large 
quantum gravity effects away from the singularity in a quantum 
theory based on this set of analytically extended  solutions. 
Large quantum fluctuations of the
metric occur even when the `classical' curvature is small.
However, since the position of the horizon also fluctuates, our 
calculation does not prove existence of large fluctuations outside
the (fluctuating) horizon.
Next, we dealt with the classical and quantum theory based on only
the $X^{\pm}>0$ region. Note that even within this region 
there is an analytic extension of the solution `above' the singularity. 
In the 
quantum theory a quantity of interest, $H$ (\ref{eq:Hr}), 
could not be represented as an operator
on the Fock space of the theory. This is unfortunate because the classical 
theory captures physically relevant collapse situations (modulo the extension 
through  the singularity). Note that ${m_R\over \kappa}$ takes the form of 
the 
usual Hamiltonian for free field theory and is also 
a quantity of interest. We do 
not know if a representation of quantum scalar field theory exists such 
that both $H$ and $m_R$ can be promoted to  operators.

For the quantum theory based on the entire $X^\pm$ plane there were no such 
difficulties. In a sense, the quantum theories on the entire plane and the 
$X^{\pm} >0$ region are unitarily inequivalent. The former uses a 
positive-negative frequency split based on the time choice 
$T={X^+ +X^-\over 2}$ 
and the latter on a time choice $t= {1\over\kappa}\ln T$.
This is very reminiscent of 
what happens in the Unruh effect in 2d \cite{unruh}, except that, there, both 
sets of modes are present. The role of the acceleration in the Unruh effect 
 is taken, here,  by 
$\kappa$. 

The following comments regarding Hawking radiation are speculative.

It seems significant that the Hawking 
temperature to leading order in the 
mass, from semiclassical calculations \cite{strominger} is independent of 
the mass and is precisely the Unruh
temperature for observers accelerating with $\kappa$. This line of thought 
has been pursued in \cite{mig} in the semiclassical context.

Calculations of the Hawking effect seem to require both right and left moving 
matter. Therefore, let us switch the right moving modes on and go back to the
quantization of \cite{jkm}. The quantum theory is a standard unitary quantum
field theory on a Fock space. But, as emphasized before, it corresponds to the
analytic extension of the usual CGHS model. We beleive that it is the analytic
extension which plays a key role in obtaining a unitary theory. A possibility 
is that the
 correlations in the quantum field which appear to have been lost by passage 
into the singularity reappear in the analytic extension beyond the singularity
in the new `universe' which lies on the `other side' of the singularity.

Note that instead of freezing the degrees of freedom corresponding to the 
right moving modes, as is done in this work, one can continue to use the 
results of \cite{jkm}, but evaluate
quantities pertaining to one sided collapse by restricting the right moving 
part of the quantum states to the (right moving) Fock vacuum. Then 
vacuum fluctuations of the right moving modes would contribute to various
quantities 
but we believe that the large 
quantum gravity effects away from the singularity (see section II.3) 
will persist. Maybe, one can also examine Hawking effect issues since the 
right moving modes are not switched off.

Finally, from the point of view of 4d quantum general relativity, we feel that
the CGHS model could be improved to a more realistic model of black holes
if somehow an internal reflecting boundary in the spacetime 
existed\cite{verlinde,fluct}. The lack of such a boundary and the fact that 
the matter is conformally coupled so that it does not `see' the singularities,
are,  
we believe,  the key unphysical features
present in the model but absent in 
(the effectively 2d) spherical collapse of a scalar field in 4d general 
relativity. The latter is, of course a physically realistic situation, but 
unfortunately technically very complicated. 
It would be interesting to try to apply the techniques of \cite{jkm}
to the model with a boundary described in \cite{verlinde,fluct} and to try to
compute  
the metric quantum fluctuations and compare results 
with those in \cite{fluct}. Since the 
boundary in \cite{fluct} is itself 
dynamically determined, it is not clear to us 
to what extent the model is solvable. 
It would be good to have a technically solvable model which was closer to 
4d collapse situations.

\vspace{3mm}

\noindent{\bf Acknowledgements:} I thank A. Ashtekar for suggesting the issue 
of large quantum fluctuations in the context of the CGHS model and for useful
discussions.  I thank S. Bose for his comments, as well as for drawing
my attention to \cite{fluct}. I thank R. Nityananda and J. Samuel for 
useful discussions. I thank P.A. Pramod for creating the figure.

\newpage

Figure 1: The black hole spacetime is embedded in it's analytic continuation
to the entire Minkowskian plane. The curly line denotes the singularity in 
the black hole spacetime and the shaded region, the (left moving) matter.

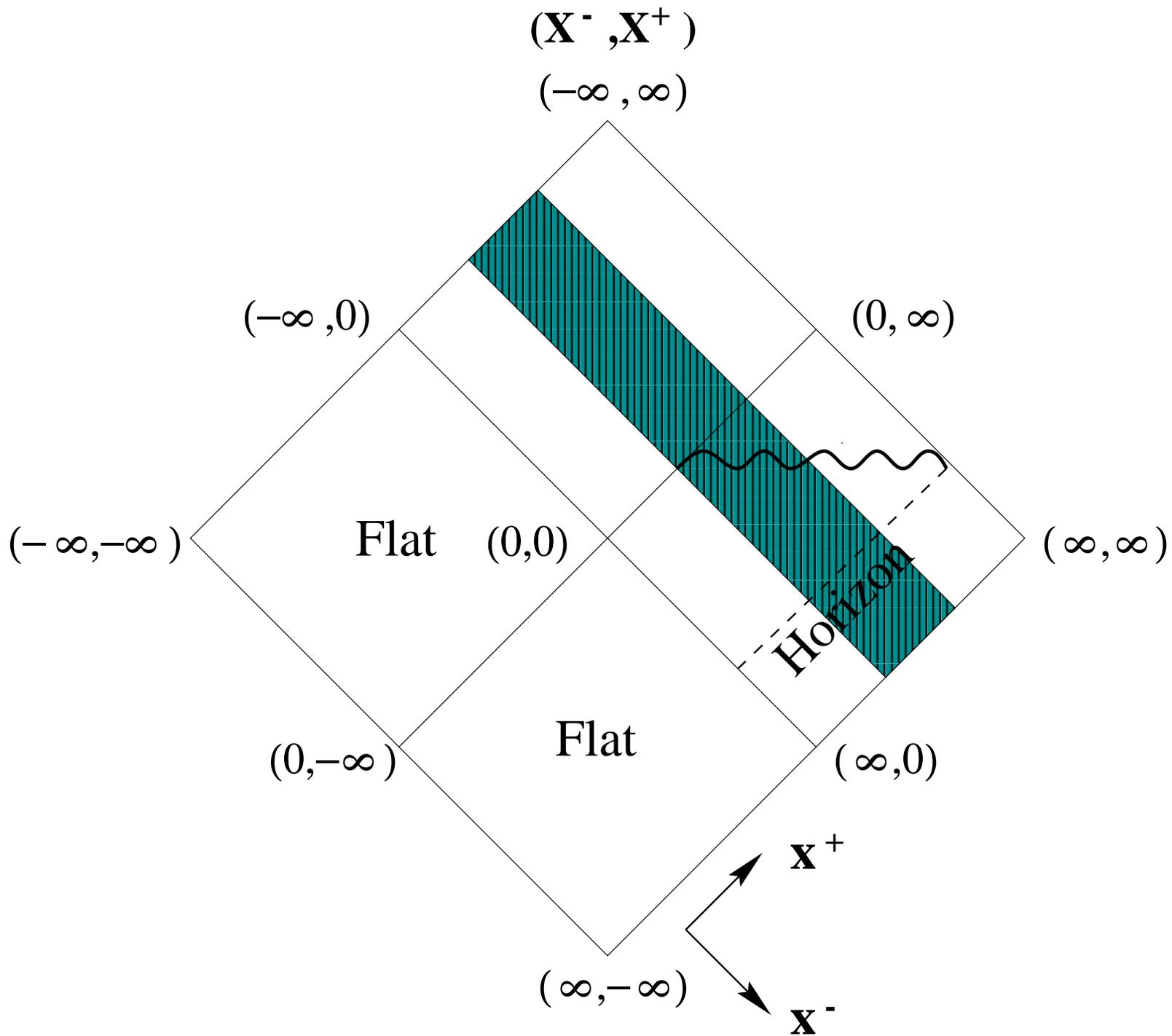
\end{document}